\documentclass[aps,prb,preprint,grouppedaddress,showpacs,showkeys]{revtex4}

\usepackage{graphicx}%
\usepackage{dcolumn}
\usepackage{amsmath}
\usepackage{multirow}
\usepackage{epsfig}
\usepackage{amssymb}
\usepackage{pifont}
\usepackage{color}

\begin{document}

\title{Linear-response calculation of the effective Coulomb interaction between closed-shell localized electrons: Cu, Zn, and ZnO}

\author{Woo-Jin Lee}
\author{Yong-Sung Kim}
\email{yongsung.kim@kriss.re.kr}
\affiliation{Korea Research Institute of Standards and Science, Yuseong, Daejeon 305-340, Korea}

\date{\today}

\begin{abstract}
A linear-response method to calculate the effective Coulomb interaction ($U$) between closed-shell localized electrons is suggested and applied to the $3d$ closed-shell systems (Cu, Zn, and ZnO) based on plane-wave basis density-functional theory calculations.
Since the closed-shell localized states are far below the Fermi level, large local perturbation potential ($\alpha$) projected to the localized states is applied to induce purposeful density response ($\Delta n$).
From the $\alpha$, the perturbation potential cost for density response onset, by which the $\Delta n$ begins to be induced, is removed.
The main screening channel for the effective Coulomb interaction is the itinerant electrons deoccupied from the perturbed localized states.
The Cu, Zn, and ZnO $3d$ electron binding energies are calculated based on the local density approximation plus $U$ with the $U$ values calculated from the linear-response, which are found to be in good agreement with experiments.
\end{abstract}


\maketitle

\section{Introduction}
The effective Coulomb interaction between the localized electrons in closed-shell systems can be important, even though their levels are located far below the Fermi level. As an example, by hybridization of the closed-shell states with some dispersed bands, in which the Fermi level lies, the correlated localized electrons can change the electronic structure near the Fermi level.
In conventional local-density-approximation (LDA) calculations, the full-occupied Zn-$3d$ states of ZnO are located within the dispersed O-$2p$ bands, by which the unrealistic strong $p$-$d$ hybridization affects the valence band states with changing the order of states at the top of the valence band and reducing the band gap.\cite{MA95,DO04,JA06,PA08,ER06,KA06} To avoid the scant description of the correlation effect in LDA, the LDA+$U$ approach\cite{LDAU,AN93} has been widely adopted for ZnO.\cite{DO04,HU06,JA06,PA08,ER06,KH09,LA05,LA08,LE07,IU08} The effective Coulomb interaction parameter $U$, however, has been chosen to fit experimental data\cite{DO04,LA08,LA05,KA06} or fit other calculations\cite{HU06,PA08,ER06} or based on atomic calculations\cite{JA06}, and spreads in a wide range of $2\sim13$ eV (Table \ref{table1}).

Constrained-density-functional theory (CDFT) calculations with the constraint of the localized state occupation ($n$) can be used for obtaining the effective Coulomb interaction parameter $U$.\cite{DE84,MC88,GU89,HY89,AN91,SO94,PI98,CO05}
The $U$ value is defined as the second derivative ($\partial^2E/\partial n^2$) of the ground state energy $E(n)$ of the constrained system or equivalently the level change ($\partial\epsilon/\partial n$) with respect to the occupation.\cite{HE66,JA78}
The LDA+$U$ can give the correct (exact DFT) ground state of the system with {\em appropriate} $U$ values through piecewise correction of the LDA total energy curve with respect to the electron number $n$.\cite{SO94,CO05}
Since there is arbitrariness in definition of the localized state occupation $n$ depending on the projection manifold,\cite{PI98,CO05,NA06}
the CDFT calculations should be performed consistently with the LDA+$U$ DFT aimed for with the same definition of $n$ and the same exchange-correlation functional to be corrected.\cite{PI98,CO05,KU06}

By employing a Lagrange multiplier ($\alpha$) in the total-energy variational form, the localized state occupation $n$ can be constrained basis-independently.\cite{PI98,CO05}
The $\alpha$ is nothing but the local perturbation potential applied to the localized states, and
the inverse linear response function $\chi^{-1}$=$\partial \alpha/\partial n$ can be used for obtaining the effective Coulomb interaction parameter $U$.
While the linear response CDFT method has been widely applied to open-shell systems,\cite{CO05,HU06,KU06,SA08,SO05,AR06,NA06} there pose some ambiguities for closed-shell systems.\cite{JA06,HU06}
Since the closed-shell states are far below the Fermi level, a small local perturbation potential $\alpha$ does not give a meaningful density change ($\Delta n$) in the localized states.
Dividing by the negligible $\Delta n$ outputs a very large inverse linear density response function $\chi^{-1}$, which seems to overestimate the effective Coulomb interaction between the closed-shell localized electrons.\cite{HU06}

In this paper, a method to calculate the effective Coulomb interaction between closed-shell localized electrons through the linear response CDFT calculations is suggested.
Small (positive) local perturbation potential $\alpha$ projected to the localized states gives only a level shift upward with little occupation change.
By applying large $\alpha$, we can induce significant localized-state density change.
From the large $\alpha$, it is needed to {\em eliminate the perturbation potential cost for density response onset} ($\epsilon_{3d}$), by which $\Delta n$ begins to be induced, in order to extract only the Hubbard interaction contribution from the $\alpha$.
The $U$ renormalization is mainly through the itinerant electrons deoccupied from the perturbed localized states.
The Cu, Zn, and ZnO $3d$ electron binding energies are calculated based on the LDA+$U$ DFT with the calculated Hubbard $U$ values.

\section{Calculational Method}
Our linear response CDFT calculations were performed based on LDA,\cite{CE80} using the
plane-wave-basis {\tt QUANTUM-ESPRESSO}\cite{Q-ES} code with ultrasoft pseudopotentials.\cite{VA90}
Local perturbation potential $\alpha$ projected to the $3d$ states was applied to a transition metal atom in the fcc Cu (4-, 32-, and 256-atom), hcp Zn (2-, 16-, and 128-atom), and wurtzite ZnO (4-, 32-, and 256-atom) supercells. The projector operator is constructed from the normalized $3d$ atomic orbitals centered at the perturbed site. Since we used the ultrasoft pseudopotentials, all scalar products between the crystal and atomic pseudo-wave-functions include the usual $S$ matrix for orthogonality.\cite{CO05,VA90}
The kinetic energy cutoff for the plane-wave basis expansion was 30 Ry.
The $2\times2\times2$ {\bf k}-point mesh including the $\Gamma$ point for the largest-size supercell and the equivalent for the smaller-sizes were used.
Experimental lattice constants of $a$=3.61 \AA~for fcc Cu, $a$=2.66 \AA~and $c/a$=1.861 for hcp Zn, and $a$=3.25 \AA~and $c/a$=1.602 for ZnO were adopted.\cite{KI91,DE98}
Here, we confine our discussions only to the rotationally invariant and site-averaged mean value of $U$, which is a good approximation for closed-shell (isotropic) bulk (homogeneous) systems.

\section{Results}
\subsection{Density Response Characteristics}
\begin{figure}[t]
\includegraphics[width=0.90\linewidth]{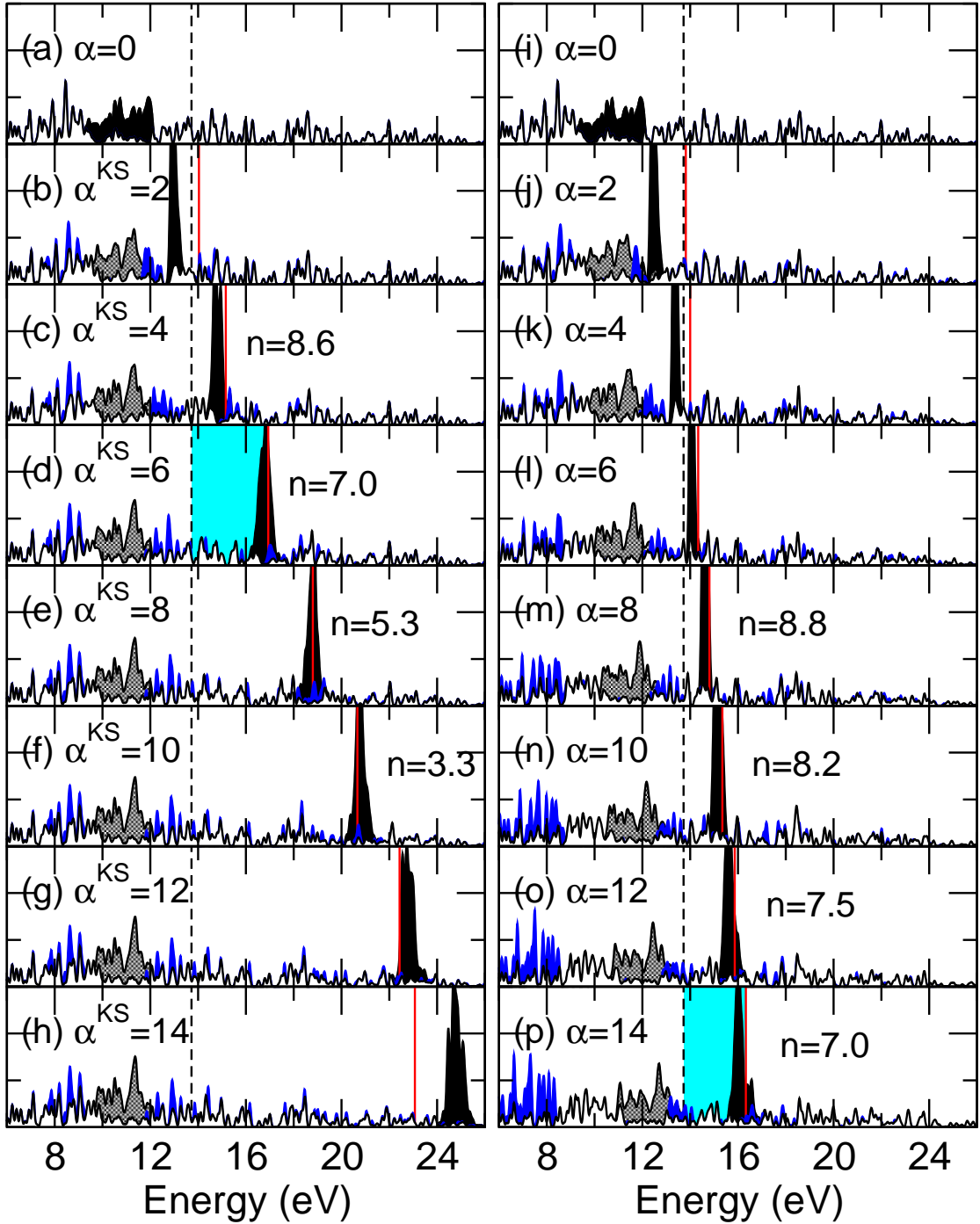}
\caption{(Color online) [(a)-(h)] Unscreened PDOS to the perturbation potential $\alpha^{\rm KS}$ (in eV) and [(i)-(p)] screened PDOS to $\alpha$ (in eV) for 4-atom-cell fcc Cu. Black-filled lines are the perturbed $3d$ states, and gray-filled lines are unperturbed $3d$ states. Blue-filled lines are the $4s$ states at the perturbed Cu site, and white-filled lines are $4s$ states at an unperturbed site. The dashed vertical lines are the Fermi level of the unperturbed system, and the solid (red) vertical lines are the Fermi level of the perturbed system. The $4s$ PDOS are multiplied by 10 for clarity.
} \label{fig1}
\end{figure}

Figure \ref{fig1} shows the calculated partial density-of-states (PDOS) of the perturbed 4-atom-cell fcc Cu. We first apply the perturbation potential $\alpha^{\rm KS}$ projected to the $3d$ states of a Cu atom in the supercell, which induces the $3d$ level shift {\em without} any screening [Fig. \ref{fig1}(a)-(h)]. The elimination of the screening to the perturbation potential was done by the non-self-consistent calculations; the perturbed DFT Hamiltonian (constructed with the $\alpha^{\rm KS}$ perturbation potential and the unperturbed charge density) was diagonalized without updating the charge density.
It can be seen that the perturbed $3d$ states move upward in energy level as the $\alpha^{\rm KS}$ increases, while the other states (the unperturbed $3d$ and $4s$ states) are barely changed with respect to the unperturbed system.
When $\alpha^{\rm KS}>3$ eV, the $3d$ states are deoccupied significantly. The electrons deoccupied from the $3d$ states fill the host electronic states. Due to the finite size of the supercell (the limited number of the host Fermi states), the Fermi level increases as the $3d$ states are deoccupied (band filling effect), which is accented by the color filling between the unperturbed and the perturbed Fermi levels in Fig. \ref{fig1}(d).
Upon the $\Delta n$, the $\alpha^{\rm KS}$ is composed of the shift ($\epsilon_{3d}^{\rm KS}$) of the $3d$ levels to the Fermi level and the Fermi level increase ($\Delta\epsilon_f^{\rm KS}$) by the band filling of the $\Delta n$ electrons.

In Fig. \ref{fig1}(i)-(p), we show the PDOS of the perturbed 4-atom-cell fcc Cu with the perturbation potential $\alpha$, which is fully screened. Since the $4s$ electrons deoccupied from the $3d$ states screen effectively the perturbation potential $\alpha^{\rm KS}$, significant deoccupation of the $3d$ states occurs only when the $\alpha$ is larger than 8 eV.
For example, the $3d^7$ configuration is found at $\alpha$=14 eV, while it is found at $\alpha^{\rm KS}$=6 eV, as shown in Fig. \ref{fig1}(d) and (p).
In the screened case, the increase of the Fermi level also occurs by the finite size of the supercell, and the screening changes the host valence states
(the $4s$ states both at the perturbed and unperturbed sites and the $3d$ states of the unperturbed Cu atom)
[see Fig. \ref{fig1}(i)-(p)].
In the process of $3d^{10}\rightarrow3d^7+(4s^*)^3$ (the $4s^*$ denotes the host electronic states changed by the screening.), the $\alpha$ contains (i) the shift ($\epsilon_{3d}$) of the $3d$ levels to the Fermi level, (ii) the Fermi level increase ($\Delta\epsilon_f$), and (iii) the Hubbard interaction contribution.
The screening in $\epsilon_{3d}$ is not related to the Hubbard interaction; when $\alpha<\epsilon_{3d}$, the host valence states are barely changed ($4s\approx4s^*$) with the $3d$ states full occupied below the Fermi level [Fig. \ref{fig1}(i)-(k)].
The $4s\rightarrow4s^*$ change of the host electronic structure accompanied by the $\Delta n$ contains the essential information of the correlation effect in $U$ renormalization.
By using a large-enough-size supercell, both the $\Delta\epsilon_f^{\rm KS}$ and $\Delta\epsilon_f$ vanish, and then the Hubbard $U$ can be evaluated from
\begin{equation}
U=\frac{(\alpha-\epsilon_{3d})-(\alpha^{\rm KS}-\epsilon_{3d}^{\rm KS})}{\Delta n},
\label{eq1}
\end{equation}
which can be interpreted as only the screening-induced level shift [$(\alpha-\epsilon_{3d})-(\alpha^{\rm KS}-\epsilon_{3d}^{\rm KS})$] by the same number of $\Delta n$, excluding the potential costs ($\epsilon_{3d}^{\rm KS}$ and $\epsilon_{3d}$) to move the perturbed states near to the Fermi level.

\subsection{Density Response Onsets}
\begin{figure}[t]
\includegraphics[width=0.70\linewidth]{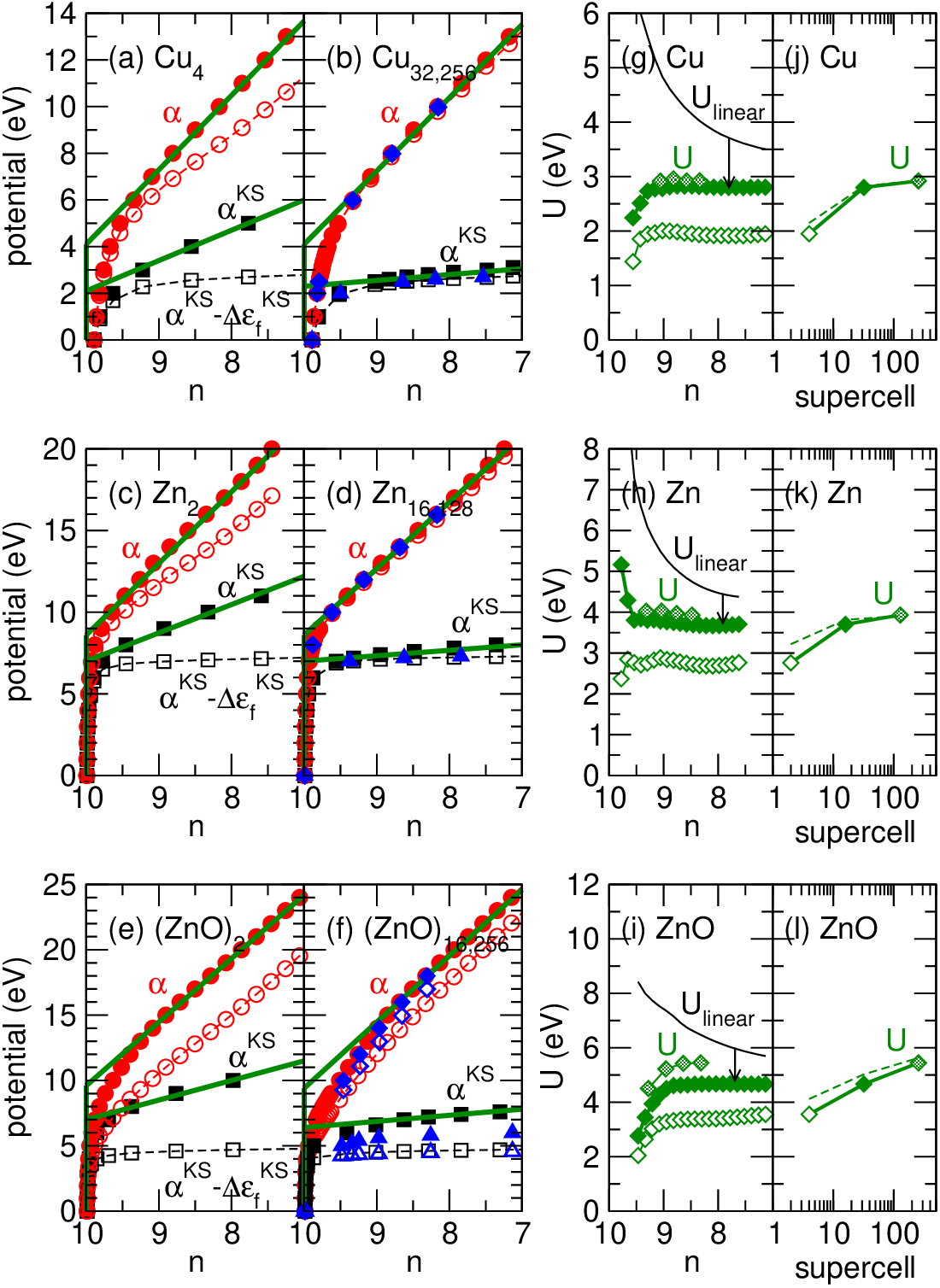}
\caption{(Color online) [(a)-(f)] Perturbation potential $\alpha$ (filled circle and diamond) and $\alpha^{\rm KS}$ (filled square and triangle) with respect to the $3d$ occupation $n$. The perturbation potentials extracted by the Fermi level increase: $\alpha-\Delta\epsilon_f$ and $\alpha^{\rm KS}-\Delta\epsilon_f^{\rm KS}$ (open symbols) are also shown. The diamond and triangle symbols are for the largest-size supercells. The (green) thick solid lines are the linear extrapolations (details are in the text).
[(g)-(i)] Calculated $U$ values with increasing $\Delta n$ for the small- (open symbols), medium- (filled symbols), and large-size (half-filled symbols) supercells of Cu, Zn, and ZnO.
The $U$ without considering the potential onsets ($\epsilon_{3d}^{\rm KS}$ and $\epsilon_{3d}$) are drawn with the (black) solid lines.
[(j)-(l)] Convergence of the $U$ values with respect to the supercell sizes for Cu, Zn, and ZnO. Including correction of the Fermi level increases ($\Delta\epsilon_f^{\rm KS}$ and $\Delta\epsilon_f$) is shown by the dashed lines.
} \label{fig2}
\end{figure}

The perturbation potential $\alpha^{\rm KS}$ and $\alpha$ with respect to the calculated $3d$ occupation $n$ are shown in Fig. \ref{fig2}(a) and (b) for the 4-, 32-, and 256-atom-cell fcc Cu.
The $\epsilon_{3d}^{\rm KS}$ and $\epsilon_{3d}$ are calculated from the extrapolation onto $n=10$ of the linear fit of the $\alpha^{\rm KS}$ and $\alpha$ with respect to $n$ in large $\Delta n$ [in the linear region of the curves in Fig. \ref{fig2}(a) and (b)].
The $\epsilon_{3d}^{\rm KS}$ and $\epsilon_{3d}$ are the potentials for the $3d$ density response onsets without and with screening, respectively (with no Hubbard interaction without $\Delta n$).
The calculated potential onsets are $\epsilon_{3d}^{\rm KS}$=2.1 eV and $\epsilon_{3d}$=4.1 eV (Table \ref{table2}), and they are almost independent of the supercell sizes, as shown in Fig. \ref{fig2}(a) and (b).
Similar calculations were done for hcp Zn (2-, 16-, and 128-atom-cell) shown in Fig. \ref{fig2}(c) and (d), and wurtzite ZnO (4-, 32-, 256-atom-cell) shown in Fig. \ref{fig2}(e) and (f).
With large $\Delta n$, the density responses are linear-like, while, near $n$=10, the perturbation potential gives little density change showing non-linear behaviors in the density-responses.
We fit linearly the density responses in the linear region (large $\Delta n$) as in the case of Cu, and calculate the perturbation potential onsets $\epsilon_{3d}^{\rm KS}$ and $\epsilon_{3d}$, listed in Table \ref{table2}.
The $3d$ states of Zn and ZnO are much deeper in energy level than those of Cu; $\epsilon_{3d}^{\rm KS}$=7.0 eV and $\epsilon_{3d}$=8.6 eV for Zn, and the $\epsilon_{3d}^{\rm KS}$=6.4 eV and $\epsilon_{3d}$=9.4 eV for ZnO.

The Fermi level increases ($\Delta\epsilon_f^{\rm KS}$ and $\Delta\epsilon_f$) are significant in small-size supercells. The large $\alpha^{\rm KS}$ increases (mainly due to $\Delta\epsilon_f^{\rm KS}$) in the smallest-size-supercell calculations [Fig. \ref{fig2}(a), (c) and (e)] are suppressed in the larger-size-supercell calculations [Fig. \ref{fig2}(b), (d), and (f)].
With increasing the supercell size, the $\alpha^{\rm KS}$ approaches $\alpha^{\rm KS}-\Delta\epsilon_f^{\rm KS}$, in which the Fermi level increase ($\Delta\epsilon_f^{\rm KS}$) is corrected, i.e., the $\Delta\epsilon_f^{\rm KS}$ goes to zero.
Then, with the large-size supercells, the density response with respect to $\alpha^{\rm KS}$ becomes very sharp near the potential onset of $\epsilon_{3d}^{\rm KS}$: 2.1 eV for Cu [Fig. \ref{fig2}(b)], 7.0 eV for Zn [Fig. \ref{fig2}(d)], and 6.4 eV for ZnO [Fig. \ref{fig2}(f)].
As the supercell size increases, the $\alpha$ also approaches $\alpha-\Delta\epsilon_f$, as shown in Fig. \ref{fig2}(a)-(b) for Cu, Fig. \ref{fig2}(c)-(d) for Zn, and Fig. \ref{fig2}(e)-(f) for ZnO.

\subsection{Hubbard U Calculations}
The Hubbard $U$ values are calculated for Cu, Zn, and ZnO by using Eq. \ref{eq1}, and shown in Fig. \ref{fig2}(g), (h), and (i), respectively.
We compare them with the $U$ values calculated through the simple linear response approach ($U_{\rm linear}=\partial\alpha/\partial n-\partial\alpha^{\rm KS}/\partial n$), which has been used for open-shell systems, without considering the potential onsets ($\epsilon_{3d}^{\rm KS}$ and $\epsilon_{3d}$).\cite{CO05}
Without considering the potential onsets, the calculated $U$ values are very large [black solid lines in Fig. \ref{fig2}(g)-(i)] and strongly depend on the $3d$ occupation $n$. This behavior originates from the non-linearity in the density responses of the $3d$ closed-shell states [Fig. \ref{fig1}(a)-(f) near $n$=10].
With extracting the potential onsets of $\epsilon_{3d}^{\rm KS}$ and $\epsilon_{3d}$ from the perturbation potentials $\alpha^{\rm KS}$ and $\alpha$, respectively, the density response behavior of the $3d$ closed-shell states becomes linear-like, and then the calculated $U$ values become rather a constant with the significant deoccupation $\Delta n$ of the $3d$ states [green diamond symbols in Fig. \ref{fig2}(g)-(i)].
The deviation from the constant $U$ value near $n$=10 is due to the deviation of the density response from the linear fit near $n$=10 [Fig. \ref{fig2}(a)-(f)].

The calculated Hubbard $U$ values with respect to the supercell sizes are shown in Fig. \ref{fig2}(j), (k), and (l) for Cu, Zn, and ZnO, respectively.
In the small-size supercells, the Hubbard $U$ is underestimated for all the three materials, similarly to the other calculations for open-shell systems,\cite{CO05} but it is fast converged when we use the 32-atom supercell for Cu [Fig. \ref{fig2}(j)] and 16-atom supercell for Zn [Fig. \ref{fig2}(k)]. In the case of ZnO, it seems to converge in the 256-atom supercell or larger [Fig. \ref{fig2}(l)].
ZnO has a very small number of density of conduction states, which are only consisted of the highly dispersed Zn-$4s$-like states near the conduction band edge.\cite{ER06,PA08,JA06} The Hubbard $U$ values calculated for the largest-size supercells are listed in Table \ref{table2}. They are 2.9 eV for Cu, 3.9 eV for Zn, and 5.4 eV for ZnO.

If we apply corrections for the Fermi level increases due to the finite size supercells by using $\alpha^{\rm KS}-\Delta\epsilon_f^{\rm KS}$ and $\alpha-\Delta\epsilon_f$ rather than $\alpha^{\rm KS}$ and $\alpha$, respectively, the convergence of the calculated Hubbard $U$ values with respect to the supercell sizes can be slightly faster, as shown by the dashed lines in Fig. \ref{fig2}(j)-(l). However, even when we apply the corrections, the convergence seems to be achieved in the large-enough-size supercells, i.e, 32- (Cu), 16- (Zn), and 256-atom (ZnO) supercells. With the large-enough-size supercells, the calculated Hubbard $U$ values are almost irrespective of whether we apply the Fermi level corrections or not, as shown in Fig. \ref{fig2}(j)-(l).

\subsection{Screening Channels}
We should address an effect of the shift ($\epsilon_{3d}$) of the $3d$ closed-shell states up to the Fermi level by applying the large $\alpha$. In this process, the host electronic states are not significantly changed ($4s\approx4s^*$) with the perturbed $3d$ states still full occupied below the Fermi level, as described above and shown in Fig. \ref{fig1}(i)-(k).
However, the perturbed $3d$ states themselves are modified. Mostly, with the $3d$ level shift up to the Fermi level, the $3d$ band width is significantly narrowed [see Fig. \ref{fig1}(i)-(p)].
It is obvious because of the weaker hybridization of the perturbed $3d$ states with the surroundings (non-resonant inter-site $d$-$d$ coupling).
By the weaker hybridization, the perturbed $3d$ states become more atomic-like embedded in the host. In tight-binding language, the hopping term is suppressed.
This is a similar situation with the traditional CDFT to calculate the $U$ parameters for the Hubbard or Anderson model, in which the hopping terms are explicitly included in the model Hamiltonian.\cite{MC88,GU89,HY89,AN91}
This isolation of the perturbed $3d$ states occurs automatically accompanied with the $3d$ level shift.
In applying the $U$ to LDA+$U$ DFT calculations, the isolation is not adequate, but the dispersed $3d$ states should be preserved for the exact ground state calculation.\cite{SO94,PI98,CO05}

In our approach, upon the density response, the $\Delta n$ holes are in the perturbed localized $3d$ states, while the $\Delta n$ electrons are in the itinerant $4sp$ states.
On the other hand, if we considered the real ground $3d$ states, the $\Delta n$ holes are in the dispersed deep $3d$ states, while the $\Delta n$ electrons are in the itinerant $4sp$ states.
The $\Delta n$ electrons in the itinerant $4sp$ states are the same for the both cases, but the $\Delta n$ holes are different in their states for the two cases.
The $\alpha-\epsilon_{\rm 3d}$ ($\epsilon_{\rm 3d}$ is a constant) with respect to the $\Delta n$ is important for the $U$ evaluation.
The $\alpha$, the level shift of the perturbed $3d$ states, upon $\Delta n$ is affected by (i) the itinerant $4sp$ electrons (screening the $\alpha$), and also by (ii) the $\Delta n$ holes in the perturbed $3d$ states (relaxation of the perturbed $3d$ orbitals by the $\Delta n$ hole generation).
The former contribution of the screening by itinerant electrons has been well known to give the main effect on the $U$ renormalization (which is also according to the Herring's indication),\cite{PI98,AN91,HE66} and the relaxation of the perturbed $3d$ orbitals upon $\Delta n$ gives a small change in $\alpha$ as a secondary contribution.
Therefore, the calculated $U$ values are correct within the first order of screening.
%
That the calculated $3d$ electron binding energies are well close to the experiments, which is discussed in the next paragraph, indicates again the screening by the itinerant electrons is the main effect on the $U$ renormalization.

\subsection{3d Electron Binding Energies}
\begin{figure}[t]
\includegraphics[width=0.70\linewidth]{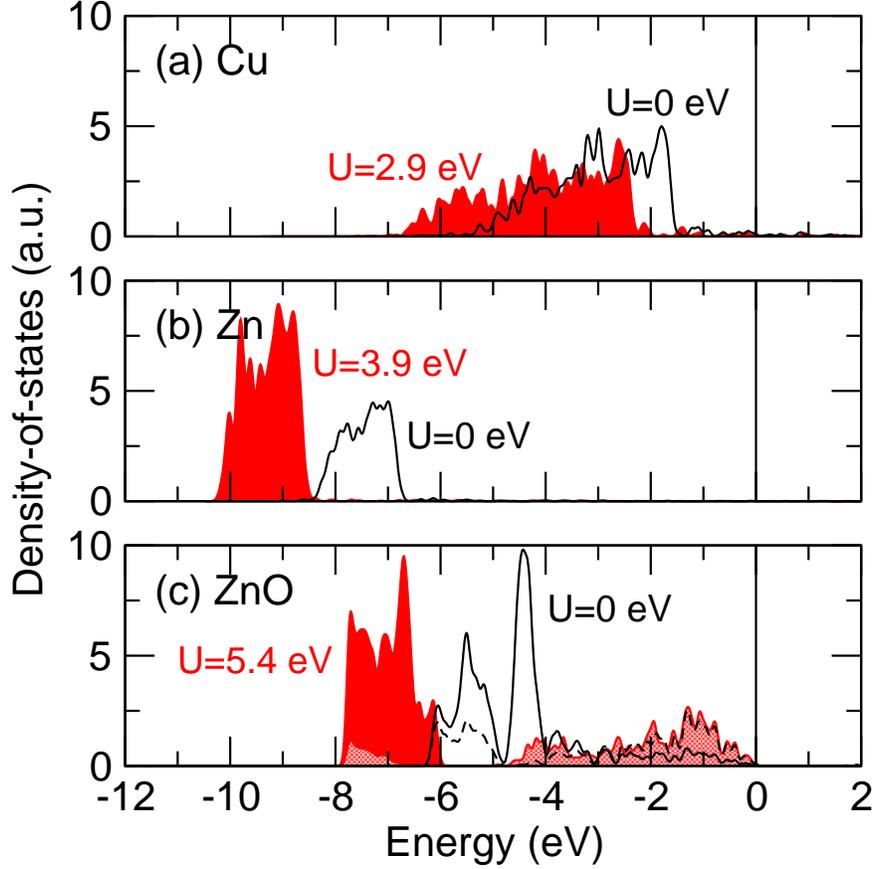}
\caption{(Color online) PDOS of bulk (a) Cu, (b) Zn, and (c) ZnO calculated by the LDA+$U$ DFT with the calculated $U$ values and by the conventional LDA.
The $3d$ states in LDA+$U$ are shown by (red) full-filled areas, and those in LDA are plotted by (black) solid lines. In (c), for ZnO, the O-$2p$ states in LDA+$U$ are shown by (red) half-filled area, and those in LDA are given by (black) dashed lines. The Fermi level is set to 0 eV in (a) Cu and (b) Zn, and the valence band maximum is set to 0 eV in (c) ZnO.
} \label{fig3}
\end{figure}

We performed the LDA+$U$ DFT calculations for bulk Cu, Zn, and ZnO with the calculated Hubbard $U$ values.
The calculated PDOS in the LDA+$U$ are shown in Fig. \ref{fig3} in comparison with the LDA results. For Cu, the LDA+$U$ ($U$=2.9 eV) yields a stronger $3d$ electron binding energy by about 0.8 eV than the LDA, as shown in Fig. \ref{fig3}(a). The experimental X-ray photo-emission spectroscopy (XPS) energy of the $3d$ state in bulk Cu is 3.1 eV,\cite{AN77} which is closer to the LDA+$U$ result.
For Zn, the LDA+$U$ calculation ($U$=3.9 eV) gives a stronger $3d$ electron binding by about 1.8 eV than the LDA, as shown in Fig. \ref{fig3}(b). The Zn $3d$ bands are located in the range of 8.5-10.2 eV below the Fermi level, and the experimental XPS energy of the $3d$ state in bulk Zn is 9.9 eV.\cite{AN77} The Zn $3d$ bands in the range of 6.8-8.5 eV in the LDA are far from the experimental value.
For ZnO, the LDA+$U$ calculation ($U$=5.4 eV) lowers the $3d$ states by about 2 eV, and the energy gap of about 1.5 eV between the Zn-$3d$ and O-$2p$ bands emerges, as shown in Fig. \ref{fig3}(c), where the Zn-$3d$ and O-$2p$ states are overlapped in energy level in the LDA calculation.
The Zn-$3d$ states are located in the range of 6.0-8.0 eV below the valence band maximum in the LDA+$U$, which is in good agreement with the XPS data of 7 eV.\cite{DO04}

\section{Summary}
In summary, we suggest a linear-response method to calculate the effective Coulomb interaction between closed-shell localized electrons. In order to apply the linear-response CDFT to closed-shell systems, we applied large perturbation potential $\alpha$
and eliminated the perturbation potential cost for the density response onset, and then the density responses become linear-like.
The main screening channel for the $U$ renormalization is the itinerant electrons deoccupied from the perturbed states.
The internally consistent LDA+$U$ DFT calculations for bulk Cu, Zn, and ZnO with the $U$ values calculated from the linear-response CDFT give the $3d$ electron binding energies well close to the experiments.

\begin{acknowledgments}
This work is supported by Nano R\&D program through the National Research Foundation of Korea funded by the Ministry of Education, Science, and Technology (No. 2009-0082489).
\end{acknowledgments}

\newpage

\begin{table}[t]
\begin{ruledtabular}
\begin{tabular}{lcc}
Authors                                     & $U$ for ZnO (eV) & Method          \\ \hline
Hu {\it et al.} [\onlinecite{HU06}]         & 2        & fit $GW$ calc.   \\
Paudel {\it et al.} [\onlinecite{PA08}]     & 3.4      & fit $GW$ calc.   \\
Janotti {\it et al.} [\onlinecite{JA06}]    & 4.7      & atomic calc.  \\
Khalid {\it et al.} [\onlinecite{KH09}]     & 5.7      & fit lattice constants \\
Dong {\it et al.} [\onlinecite{DO04}]       & 6        & fit PES expt. \\
Lathiotakis {\it et al.} [\onlinecite{LA08}]& 6.5      & fit PES expt. \\
Lany {\it et al.} [\onlinecite{LA05}]       & 7        & fit PES expt. \\
Erhart {\it et al.} [\onlinecite{ER06}]     & 7.5$^*$  & fit SIC calc. \\
Lee {\it et al.} [\onlinecite{LE07}]        & 8.5      & fit PES expt. \\
I\c usan {\it et al.} [\onlinecite{IU08}]   & 9        & -     \\
Karazhanov {\it et al.} [\onlinecite{KA06}] & 11.1     & CDFT calc.$^\dagger$ \\
Karazhanov {\it et al.} [\onlinecite{KA06}] & 13.0     & fit PES expt. \\
This work                                   & 5.4      & linear response CDFT \\
\end{tabular}
\end{ruledtabular}
\caption{Comparison of the Hubbard $U$ parameters used in previous LDA+$U$ calculations for Zn-$3d$ states in ZnO. $^*$This value is $U-J$. SIC denotes the self-interaction-correction, and PES is the photo-emission spectroscopy. $^\dagger$Details of the CDFT calculation are not described in the reference; if ionized charges from the localized states do not reside in the system or if the nonlinearity in density response is not corrected, a large $U$ value can be obtained.
}\label{table1}
\end{table}

\begin{table}[b]
\begin{ruledtabular}
\begin{tabular}{ccccc}
Material & $\epsilon_{3d}^{\rm KS}$ (eV) & $\epsilon_{3d}$ (eV) & $U$ (eV) \\ \hline
Cu       & 2.1                           & 4.1                  & 2.9 \\
Zn       & 7.0                           & 8.6                  & 3.9 \\
ZnO      & 6.4                           & 9.4                  & 5.4 \\
\end{tabular}
\end{ruledtabular}
\caption{Calculated potential onsets, $\epsilon_{3d}^{\rm KS}$ and $\epsilon_{3d}$, and Hubbard $U$ values for fcc Cu, hcp Zn, and wurtzite ZnO are listed. All values are from the largest-size supercells in our calculations.
}\label{table2}
\end{table}

\newpage

\end{document}